\documentclass[%
 reprint,
nofootinbib,
 amsmath,amssymb,
 aps,
prl,
floatfix
]{revtex4-1}

\usepackage{lmodern}
\usepackage{graphicx}
\usepackage{bm}
\usepackage{hyperref}
\hypersetup{linktocpage,colorlinks,citecolor={blue},pdfdisplaydoctitle=true,pdfpagemode=UseOutlines,bookmarksnumbered=true}
\usepackage{mathrsfs,dsfont}
\usepackage{color}
\usepackage{cleveref}

\newcommand{\bra}[1]{\langle #1 |} 
\newcommand{\ket}[1]{| #1 \rangle } 
\newcommand{\upd}{\mathrm{d}}
\newcommand{\tr}{\mathrm{tr}}
\newcommand{\ie}[0]{\textit{i.e.} }
\newcommand{\eg}[0]{\textit{e.g.} }

\newcommand{\e}[0]{\mathrm{e}}

\DeclareMathOperator*{\argmin}{argmin}

\begin{document}

\title{Variational method in relativistic quantum field theory without cutoff}

\author{Antoine Tilloy}
\email{antoine.tilloy@mpq.mpg.de}
\affiliation{Max-Planck-Institut f\"ur Quantenoptik, Hans-Kopfermann-Stra{\ss}e 1, 85748 Garching, Germany}
\affiliation{Munich Center for Quantum Science and Technology (MCQST), Schellingstr. 4, D-80799 M\"unchen}

\begin{abstract}
\noindent
The variational method is a powerful approach to solve many-body quantum problems non perturbatively. However, in the context of relativistic quantum field theory (QFT), it needs to meet 3 seemingly incompatible requirements outlined by Feynman: extensivity, computability, and lack of UV sensitivity. In practice, variational methods break one of the 3, which translates into the need to have an IR or UV cutoff. In this letter, I introduce a relativistic modification of continuous matrix product states that satisfies the 3 requirements jointly in $1+1$ dimensions. I apply it to the self-interacting scalar field, without UV cutoff and directly in the thermodynamic limit. Numerical evidence suggests the error decreases faster than any power law in the number of parameters, while the cost remains only polynomial.
\end{abstract}

\maketitle

\paragraph*{Introduction -- } Quantum field theory (QFT) lies at the root of fundamental physics, and is the most fundamental approach we so far have to understand microscopic phenomena. A vexing problem of theoretical physics is that QFTs are rarely ever solvable. We seem to know the rules of particle physics, at least to a good precision, but hardly know what they give in general.

Until recently, there were essentially two approaches to deal with QFT approximately: perturbation theory \cite{peskin1995} and lattice Monte-Carlo \cite{creutz1983,flag2014}. The first provides results without cutoff in momenta, valid ``all the way down'' for a true QFT, but accurate only for small coupling. The second works at strong coupling, but introduces a short (UV) and long (IR) distance cutoff.

Variational methods are a seducing third way. The idea is to put forward a manifold $\mathcal{M}$ of quantum states $\ket{\psi_\mathbf{w}}$, specified by a small number of parameters $\mathbf{w}$, and to minimize the expectation value of the Hamiltonian $H$ over these parameters.
\begin{equation}
    \ket{\mathsf{ground}} \simeq \ket{\psi_\mathbf{w}} \;\text{for}\; \mathbf{w} = \argmin_\mathcal{M} \frac{\bra{\psi_\mathbf{w}} H \ket{\psi_\mathbf{w}}}{\langle\psi_\mathbf{w}|\psi_\mathbf{w}\rangle}
\end{equation}
If the manifold is guessed right, this can provide a good non-perturbative approximation to the ground state, from which one may then compute observables.

In the context of relativistic QFT, the variational method was submitted to a devastating criticism by Feynman in $1987$, who listed $3$ crippling objections, or rather requirements on the state manifold that could not possibly be met jointly \cite{feynman1988}. The first is extensivity: the states should be extensive, in the sense that increasing the system size (IR cutoff) should increase the dimension of the manifold at most linearly (and not exponentially). The second is computability: the states should be such that expectation values of local observables can be computed reasonably efficiently. This is needed to minimize the energy, but also to extract physical predictions once the state is known. The last requirement is specific to relativistic QFT, and is a lack of sensitivity to UV features. Since the energy density of relativistic QFT is dominated by arbitrarily large momenta, minimizing the energy will adjust the state parameters towards fitting shorter and shorter distances, paradoxically degrading the accuracy at physically relevant length-scales. According to Feynman, only Gaussian states could fit these 3 requirements, which excluded the variational method for interacting theories.

Modern variational approaches swallow at least one of Feynman's bullets. Hamiltonian truncation (HT) and its renormalized refinements \cite{rychkov2015} use a vector space, the free Fock space, as state manifold. With an IR cutoff, energy levels get discretized, and there is only a finite number of basis states under a truncation energy $E_T$. With these two cutoffs, the energy minimization is a simple finite dimensional linear problem. On the other hand, HT clearly breaks extensivity, as the number of basis states and thus parameters grows exponentially as the system size (IR cutoff) is increased. Because the prefactors are favorable, and extrapolations reliable \cite{eliasmiro2016}, HT can still be a precise method in practice \cite{hogervorst2015,eliasmiro2017-1,eliasmiro2017-2,elias-miro2020}.

Tensor network states \cite{verstraete2004,cirac2020matrix}, defined on the lattice, are a naturally extensive class of states. In their $1+1$ dimensional incarnation, the matrix product states (MPS) \cite{fannes1992}, local observables are also efficiently computable. In 2010, Cirac and Verstraete took the continuum limit of MPS, to get the continuous MPS (CMPS) \cite{verstraete2010}. While it provides an efficient ansatz for non-relativistic QFT, it still suffers from Feynman's third objection in the relativistic context.

In practice CMPS can still be used for relativistic QFT, but one needs to add a UV cutoff in the Hamiltonian, that acts as Lagrange multiplier to prevent the state from even fitting the UV \cite{haegeman2010-relativistic,stojevic2015}. This limits the range of validity of the results, breaks the strict variational nature of the approach, and partially defeats the purpose of going to the continuum in the first place. This UV difficulty is understandable: at short distances, true $1+1$d relativistic QFT are conformal field theories, valid all the way down, without cutoff scale. CMPS cannot capture this short distance behavior by construction, as they are regular at short distances. The necessity of a UV cutoff thus seems inevitable.

This situation is frustrating because, at least for super-renormalizable and even asymptotically free theories, the UV behavior CMPS fail to capture is otherwise trivial. Can we not include this ``free'' behavior exactly? My objective in this letter is to show that in $d=1+1$ dimensions, this is possible. The requirements of Feynman can be jointly satisfied: one can put forward an extensive and efficiently computable class of states, that comes without cutoff (UV or IR), and that gives the energy density and all local observables to arbitrary precision upon optimization.

This new class of states, the relativistic CMPS (RCMPS), borrow most of their definition from CMPS. The new ingredient is a change of operator basis and Fock space (Bogoliubov transform), that provides the right large momentum behavior. This new Fock space is the Fock space adapted to the free part of the theory, which is precisely the one used in the Hamiltonian truncation approach (up to a removal of the IR cutoff). RCMPS can thus be seen as a hybridization of HT and tensor network methods.

Defining RCMPS and obtaining their basic properties is rather straightforward and done in the present letter. However, evaluating the energy expectation value for a given theory and then minimizing it by varying the state parameters, requires slightly lengthier computations. They are presented in full glory in a companion paper \cite{rcmps_article}, that also provides more in depth discussions.

\paragraph*{The model --} The prototypical theory we will apply RCMPS to is the self-interacting scalar, a.k.a. $\phi^4_2$ theory. The model is specified by its Hamiltonian
 \begin{equation}\label{eq:phi4Hamiltonian}
    H= :\left[\int_{\mathbb{R}} \frac{\pi^2}{2}+ \frac{(\partial_x \phi)^2}{2} + \frac{m^2}{2} \phi^2 + g\,  \phi^4 \,\right]:.
\end{equation}
The normal-ordering is done with respect to the operators $a_k,a_k^\dagger$ that diagonalize the free part of the Hamiltonian obtained for $g=0$. More precisely, the field operators admit the mode expansion
\begin{align}\label{eq:modeexpansion}
    \phi(x) &= \frac{1}{2\pi} \int \upd k \sqrt{\frac{1}{2 \, \omega_k}} \left(\e^{ikx} a_k + \e^{-ikx} a^\dagger_k \right) \\
        \pi(x) &= \frac{1}{2i\pi} \int \upd k \sqrt{\frac{\omega_k}{2}} \left(\e^{ikx} a_k - \e^{-ikx} a^\dagger_k \right) \, ,
\end{align}
where $\omega_k=\sqrt{m^2+k^2}$ and $[a_k,a^\dagger_{k'}]=2\pi \delta(k-k')$.
Crucially, in $1+1$ dimensions, the normal-ordering $:\diamond:$ is sufficient to renormalize all the UV divergences (corresponding to perturbation theory tadpoles), and $H$ is a legitimate self-adjoint operator \cite{glimm1987,fernandez1992}. While easy to define, $\phi^4_2$ theory is surprisingly difficult to solve. It is not integrable, and its behavior at strong coupling $g\gtrsim m^2$ is challenging to probe numerically. 

This model has been studied with a wide variety of methods: renormalized Hamiltonian truncation \cite{rychkov2015,eliasmiro2017-1,eliasmiro2017-2}, Monte-Carlo \cite{bronzin2019}, tensor network renormalization \cite{kadoh2019,delcamp2020}, matrix product states \cite{milsted2013,vanhecke2019}, resummed perturbation theory \cite{serone2018}. All these methods, apart from perturbation theory, require at least one cutoff, UV or IR, and thus extrapolations.
 
 \paragraph*{The state manifold --} A RCMPS is a quantum state belonging to the free Fock space, parameterized by 2 $D\times D$ complex matrices $Q,R$ and defined as
\begin{equation}\label{eq:def}
    \ket{Q,R} = \tr\left\{\mathcal{P} \exp\left[\int \upd x \, Q\otimes \mathds{1} + R\otimes a^\dagger(x)\right]\right\}\ket{0}_a \, .
\end{equation}
In this formula, the trace is taken over the finite $D$ dimensional auxiliary space of matrices, $\mathcal{P}\exp$ is the path-ordered exponential, and $a^\dagger(x)$ is a creation operator such that $[a(x),a^\dagger(y)]=\delta(x-y) \mathds{1}$. The state $\ket{0}_a$ is the Fock vacuum annhilitated by all the $a(x)$. The bounds in the integral can be an interval $[-L,L]$ or $\mathbb{R}$, that is directly the thermodynamic limit, which makes extensivity manifest. The \emph{bond dimension} $D$ is a proxy the expressiveness of the state manifold: the larger it is, the more knobs one can tune to lower the energy and fit the true ground state.

The definition \eqref{eq:def} would be that of a standard CMPS if $a^\dagger(x)$ were chosen to be a local creation operator associated to the canonically conjugated pair $\phi(x),\pi(x)$, \ie $\psi^\dagger(x) = \sqrt{\nu/2} \phi(x) - i(\sqrt{2\nu})^{-1}\pi(x)$ for some $\nu$. This would be a natural non-relativistic choice, preserving locality, but creating the UV difficulties we discussed previously.

Instead, as the notation suggests, I take $a(x)$ to be the Fourier transform of $a_k$
\begin{align}\label{eq:fourier}
    a(x) &= \frac{1}{2\pi}\int \upd k \, \e^{ikx} a_k \, .
\end{align}
This operator rarely ever appears in relativistic QFT. It is not a local function of the fields $\phi(x),\pi(x)$ because of the (lack of) factor $\sqrt{\omega_k}$. It also does not transform covariantly. Fortunately this property is not required here (nor would it bring anything). As discussed in the companion paper, this choice for $a(x)$ is actually not the only one, but it is arguably the simplest to make the ansatz match the UV behavior of the QFT while retaining exactly computable correlation functions.

Indeed, if $R$ and $Q$ are zero, the RCMPS is just the Fock vacuum $\ket{0}_a$, which is the ground state at $g=0$. It thus has the the right short distance behavior for free, without the need for any parameter tuning. When $R,Q$ are non-zero, the state lies in the same Fock space and the UV behavior remains unchanged. One can further prove that local expectations values, like the energy density, are finite and well behaved \cite{rcmps_article}.

\paragraph*{Computations, in a nutshell --} Since $a(x)$ verifies the same commutation relations as $\psi(x)$, the standard CMPS formulas \cite{haegeman2013}, which depend only on this algebra, can be reused in the RCMPS context. In particular, all local normal-ordered correlations functions of $a(x)$ have a compact algebraic expression as a trace over finite dimensional matrices.

This is seen by introducing the generating functional $\mathcal{Z}_{j',j}$:
\begin{equation}\label{eq:generatingfunction}
\mathcal{Z}_{j',j}=\frac{\bra{Q,R} \exp\left(\int j'\, a^\dagger \right) \exp\left(\int j\, a \right) \ket{Q,R}}{\langle Q,R|Q,R\rangle} \,,
\end{equation}
which can be used to compute all normal-ordered correlation $N$-point functions of $a(x),a^\dagger(x)$ by taking functional derivatives. One can show \cite{haegeman2013} that this generating functional has an exact expression:
\begin{equation}
    \mathcal{Z}_{j',j} \! = \! \tr\left\{\mathcal{P} \exp\!
    \bigg[\int\upd x \, \mathbb{T}  + j(x)  R \otimes \mathds{1} +j'(x) \mathds{1}\otimes \bar{R}\bigg]\right\}\,
\end{equation}
where $\mathbb{T} = Q\otimes \mathds{1} + \mathds{1}\otimes \bar{Q} + R\otimes \bar{R}$ is the transfer operator and the trace is taken over the tensor product of two copies of the original $D$ dimensional auxiliary Hilbert space. For example, on the interval $[-L,L]$, and for $L\geq x\geq y\geq -L$ this gives:
\begin{equation}\label{eq:twopointexplicit}
    \langle a^\dagger(x) a(y) \rangle \!=\! \tr\Big[ \e^{(L-x)\mathbb{T}} (\mathds{1}\otimes \bar{R})
\e^{(x-y)\mathbb{T}} (R\otimes\mathds{1}) \e^{(y+L)\mathbb{T}}\Big],
\end{equation}
and other correlation functions take a similar form. This formula can be further simplified in the thermodynamic limit by making a proper choice of gauge \cite{rcmps_article}.

We are not yet done if we want to evaluate the energy density. The latter is local in the fields $\phi,\pi$ which are not local in $a,a^\dagger$, \eg
\begin{align}
    \phi(x) &= \frac{1}{2\pi} \int \upd k \sqrt{\frac{1}{2 \, \omega_k}} \left(\e^{ikx} a_k + \e^{-ikx} a^\dagger_k \right) \nonumber \\
     &= \frac{1}{2\pi} \int  \frac{\upd k \, \upd y}{\sqrt{2 \, \omega_k}} \left(\e^{ik(x-y)} a(y) + \e^{-ik(x-y)} a^\dagger(y) \right) \nonumber\\
     &=\int \upd y\;  J(x-y) \left[a(y) +  a^\dagger(y) \right]\,
\end{align}
where $J(x)$ is a smooth kernel away from $x=0$, which, crucially, decays exponentially as $|x|\rightarrow + \infty$, with a rate proportional to $m$. Rewriting the Hamiltonian density in terms of $a(x),a^\dagger(x)$ thus yields (naively) as many integrals as the degree in the fields. This lack of strict locality is a technical inconvenience but it is not expected to make the RCMPS less good at approximating the ground state~\cite{rincon2015}.

Ultimately, expressing the Hamiltonian density $h$ as an integral of $a$'s and then using the exact expressions for expectation values of $a$'s, one can write
\begin{equation}\label{eq:fQR}
    \langle h\rangle_{Q,R} = f(Q,R)
\end{equation}
where $f$ involves nested integrals of traces of matrices. This function $f$ is finite, but takes a non-trivial form, which is derived in the companion paper \cite{rcmps_article}. What matters for us here is that $f$ can be efficiently  evaluated numerically by solving ordinary differential equations with a cost $\propto D^3$. 

All that remains is to minimize over $R$ and $Q$ to find an approximation of the ground state. In principle, one could use whatever numerical solver, \eg based on quasi-Newton methods. It turns out that for the minimization to be efficient, it is better to use a more elaborate tangent space approach \cite{vanderstraeten2019tangentspace} and implement fast approximate imaginary time evolution \cite{rcmps_article}. Again, all that matters is that it can be done efficiently, with a cost $D^3$ per iteration using backpropagation methods.

\begin{figure}
    \centering
    \includegraphics{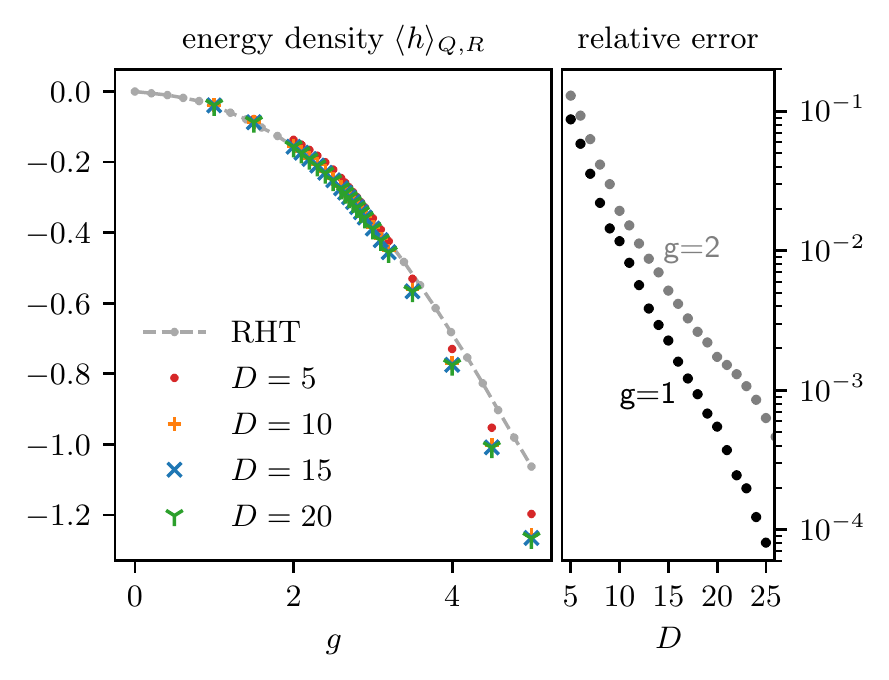}
    \caption{Left: Approximate ground state energy density as a function of the coupling $g$ for $m=1$, compared with the RHT results of \cite{rychkov2015}. Right: relative error in the energy density as a function of the bond dimension $D$, taking the results at $D=32$ for $g=1$ and $g=2$ as  close to exact comparisons.}
    \label{fig:energy}
\end{figure}

\paragraph*{Results --} The results for the ground state energy density of $\phi^4_2$ are shown in Fig. \ref{fig:energy} and compared with the renormalized Hamiltonian truncation (RHT) computations of~\cite{rychkov2015}. Even a very moderate $D=5$, corresponding to $50$ independent real parameters, provides qualitatively accurate results. For $g=1$ and $g=2$, I pushed the computations to $D=32$, to get a near exact point of comparison to evaluate the error at lower bond dimensions. This suggests that the error decreases exponentially as a function of the bond dimension.

For large values of the coupling $g\geq 3$, the RCMPS give substantially lower values for the energy density than RHT results from \cite{rychkov2015}. Since the RCMPS results are rigorous upper bounds, it means its energies are closer to the true one.

Local observables can be straightforwardly computed once the ground state is known, just like Feynman ordered, since correlation functions have an explicit form. As an illustration, one can look at the field expectation value $\langle \phi \rangle$ that shows a clear symmetry breaking around the expected value $g_c\simeq 2.77$ \cite{delcamp2020}. Results converge quickly as a function of the bond dimension, at least deep in the symmetric and symmetry broken phases. Close to the critical point, the precision is lower as expected, as entanglement diverges, and a finite entanglement scaling approach \cite{pollmann2009,stojevic2015,vanhecke2019} would be needed to enhance the precision. Using standard tangent space techniques for CMPS \cite{vanderstraeten2019tangentspace}, one could also evolve a RCMPS in real time to probe the full dynamics and compute the excitation spectrum variationally.

\begin{figure}
    \centering
    \includegraphics{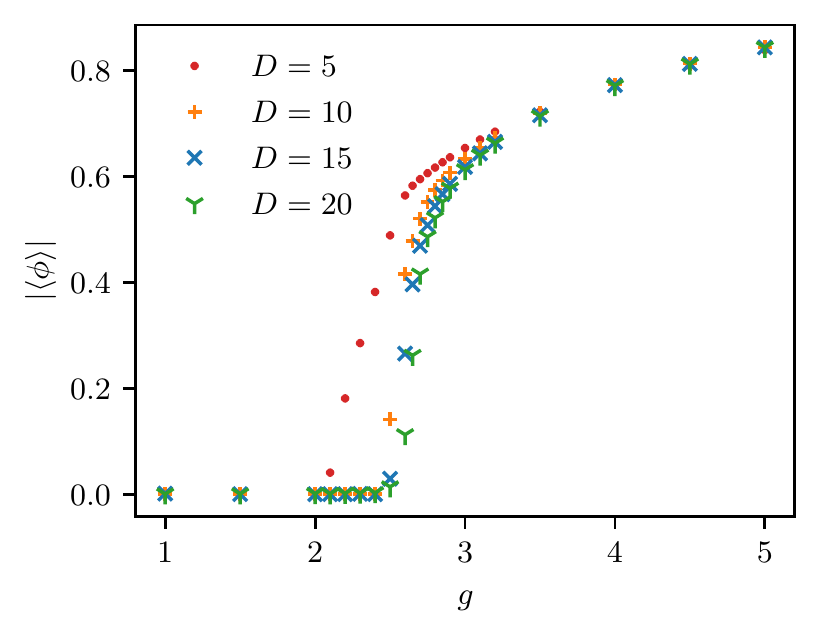}
    \caption{Field expectation value as a function of the coupling. A symmetry breaking, in the Ising universality class, is expected around $g_c\simeq 2.77$. Away from the critical point, the convergence in bond dimension is fast as expected.}
    \label{fig:correlations}
\end{figure}

\paragraph*{Discussion --} The variational method with RCMPS is efficient to solve $1+1$ dimensional relativistic QFT, without cutoff. Even for the moderate bond dimensions $D$ probed here, one gets rigorous and competitive energy upper bounds, and a hint of exponential convergence. A convergence faster than any power law is proved for gapped systems on the lattice with MPS \cite{huang2015computing}, and so is naturally expected here as well, at least away from the critical point. Adapting such a proof in the QFT context would also be a crucial advance. If the fast convergence is confirmed theoretically, and since the cost is only polynomial in $D$ ($ \propto D^3$), variational minimization with RCMPS would get as close to an exact solution as one can hope for in non-integrable relativistic QFT. 

Extending RCMPS to $2+1$ and $3+1$ dimensional relativistic QFT is currently non-trivial. The first difficulty comes from the tensor network side. A reasonable generalization of CMPS to higher dimensions, the continuous tensor network states (CTNS), was proposed in \cite{tilloy2019}. However, it does not come with an efficient toolbox to compute expectation values. Analogs of formula \eqref{eq:twopointexplicit} exist, but the trace is taken over an infinite dimensional Hilbert space, and is thus costly to even just approximate. As CMPS, CTNS are also adapted only to non-relativistic QFT and require a UV cutoff to deal with relativistic ones \cite{karanikolaou2020gaussian}. There comes the second difficulty, related to relativistic QFT themselves. In higher dimensions, normal-ordering is not sufficient, renormalization requires explicit counter terms, and the Hilbert space is no longer the free Fock space \cite{glimm1968}. Hence the Bogoliubov transform I proposed to adapt the states to the UV behavior of the free theory would no longer be sufficient. One would likely have to work with a more complicated Hilbert space.

Before these difficult questions are addressed, RCMPS are already be applicable to a wide variety of theories in $1+1$ dimension with polynomial and exponential interactions \cite{rcmps_article}.

\bibliography{main}

\end{document}